\documentclass[12pt, letterpaper]{article}
\usepackage{jheppub}
\usepackage{amsmath}
\usepackage{amssymb}
\usepackage{graphicx}
\usepackage{braket}
\usepackage{slashed}
\usepackage{color}

\newcommand{\comment}[1]{}
\newcommand{\oprod}[2]{\left| #1 \right\rangle\!\! \left\langle #2 \right| } 
\newcommand{\ev}[1]{\left\langle#1\right\rangle}
\newcommand{\be}{\begin{equation}}
\newcommand{\ee}{\end{equation}}

\newcommand{\trace}{\ensuremath{\operatorname{Tr}}}
\newcommand{\thetaorder}[1]{\ensuremath{\mbox{\ensuremath{\mathcal T}}}\left[#1\right]}
\newcommand{\calA}{\ensuremath{\mathcal{A}}}
\newcommand{\sign}{\ensuremath{\operatorname{sign}}}
\newcommand{\replicaoperator}{\ensuremath{\mathcal{D}}}
\newcommand{\dprime}{\raisebox{1.5pt}{$\scriptstyle{\prime\prime}$}}

\title{Proof of the Quantum Null Energy Condition}

 \author[a,b]{Raphael Bousso,}
\author[a,b]{Zachary Fisher,} 
\author[a,b]{Jason Koeller,} 
\author[a,b]{Stefan Leichenauer,} 
 \author[c]{and Aron C. Wall} 
\affiliation[a]{Center for Theoretical Physics and Department of Physics,\\
University of California, Berkeley, CA 94720, U.S.A.} 
\affiliation[b]{Lawrence Berkeley National Laboratory, Berkeley, CA 94720, U.S.A.} 
 \affiliation[c]{Institute for Advanced Study, Princeton, NJ 08540, USA}

\abstract{We prove the Quantum Null Energy Condition (QNEC), a lower bound on the stress tensor in terms of the second variation in a null direction of the entropy of a region. The QNEC arose previously as a consequence of the Quantum Focussing Conjecture, a proposal about quantum gravity. The QNEC itself does not involve gravity, so a proof within quantum field theory is possible. Our proof is somewhat nontrivial, suggesting that there may be alternative formulations of quantum field theory that make the QNEC more manifest.

Our proof applies to free and superrenormalizable bosonic field theories, and to any points that lie on stationary null surfaces. An example is Minkowski space, where any point $p$ and null vector $k^a$ define a null plane $N$ (a Rindler horizon). Given any codimension-2 surface $\Sigma$ that contains $p$ and lies on $N$, one can consider the von Neumann entropy $S_\text{out}$ of the quantum state restricted to one side of $\Sigma$.   A second variation $S_\text{out}^{\prime\prime}$ can be defined by deforming $\Sigma$ along $N$, in a small neighborhood of $p$ with area $\cal A$. The QNEC states that $\langle T_{kk}(p) \rangle \ge \frac{\hbar}{2\pi} \lim_{{\cal A}\to 0}S_\text{out}^{ \prime\prime}/{\cal A}$.}

\begin{document}
\maketitle


\section{Introduction}

The null energy condition (NEC) states that $T_{kk} \equiv T_{ab} k^a k^b \ge 0$, where $T_{ab}$ is the stress tensor and $k^a$ is a null vector.  This condition is satisfied by most reasonable classical matter fields. In Einstein's equation, it ensures that light-rays are focussed, never repelled, by matter. The NEC underlies the area theorems~\cite{Haw71,BouEng15a} and singularity theorems~\cite{Pen65,HawEll,Wald}, and many other results in general relativity \cite{Morris:1988tu,Friedman:1993ty,Farhi:1986ty,Tipler:1976bi,Hawking:1991nk,Olum:1998mu,Visser:1998ua,Penrose:1993ud,Gao:2000ga}.

However, quantum fields violate all local energy conditions, including the NEC \cite{Epstein:1965zza}.  The energy density $\langle T_{kk} \rangle$ at any point can be made negative, with magnitude as large as we wish, by an appropriate choice of quantum state. In a stable theory, any negative energy must be accompanied by positive energy elsewhere.  Thus, positive-definite quantities linear in the stress tensor that are bounded below may exist, but must be nonlocal.  For example, a total energy may be obtained by integrating an energy density over all of space; an ``averaged null energy'' is defined by integrating $\langle T_{kk} \rangle$ along a null geodesic \cite{Borde:1987qr,Wald:1991xn,Klinkhammer,Verch,Graham:2007va,Hofman:2009ug}.  In some field theories, ``quantum energy inequalities'' have also been shown, in which an integral of the stress-tensor need not be positive, but is bounded below \cite{Ford:1999qv}.

\begin{figure}[t]
	\centering
	\includegraphics[height=2in]{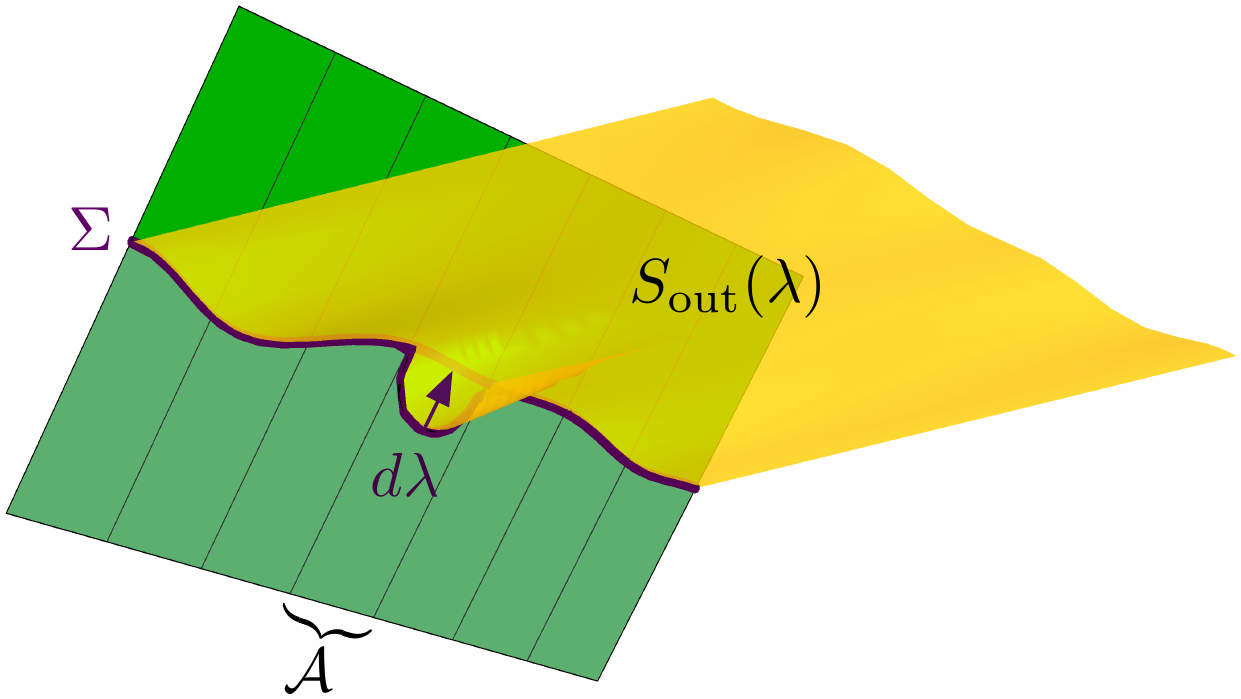}
	\caption{The spatial surface $\Sigma$ splits a Cauchy surface, one side of which is shown in yellow. The generalized entropy $S_\text{gen}$ is the area of $\Sigma$ plus the von Neumann entropy $S_\text{out}$ of the yellow region. The quantum expansion $\Theta$ at one point of $\Sigma$ is the rate at which $S_\text{gen}$ changes under a small variation $d\lambda$ of $\Sigma$, per cross-sectional area $\cal A$ of the variation. The Quantum Focussing Conjecture states that the quantum expansion cannot increase under a second variation in the same direction. If the classical expansion and shear vanish (as they do for the green null surface in the figure), the Quantum Null Energy Condition is implied as a limiting case. Our proof involves quantization on the null surface; the entropy of the state on the yellow spacelike slice is related to the entropy of the null quantized state on the future (brighter green) part of the null surface.}
	\label{fig:setup}
\end{figure}

In this article, we will consider a new type of lower bound on $\langle T_{kk} \rangle$ at a single point $p$.  Here the bound itself is computed from a nonlocal object: the von Neumann entropy $S_\text{out}[\Sigma] \equiv -\trace (\rho \ln \rho)$ of the quantum fields restricted to some finite or infinite spatial region whose boundary $\Sigma$ contains $p$, is normal to $k^a$, and has vanishing null expansion at $p$. (There are infinitely many ways of choosing such $\Sigma$ for any $(p,k^a)$.) Then a lower bound is given by the second derivative of $S_\text{out}$, under deformations of an infinitesimal area element \(\mathcal{A}\) of $\Sigma$ in the $k^a$ direction at $p$ (see Figure \ref{fig:setup}):
\begin{equation}\label{QNEC}
\langle T_{kk} \rangle \ge \frac{\hbar}{2\pi {\cal A}} S_\text{out}^{ \prime\prime}[\Sigma]~.
\end{equation}

We call \eqref{QNEC} the {\em Quantum Null Energy Condition} (QNEC)~\cite{BouFis15a}. The quantity $S_\text{out}$ is divergent but its derivatives are finite. (A more rigorous formulation in terms of functional derivatives will be given in the main text.)
Note that the right hand side can have any sign. If it is positive, then the QNEC is stronger than the NEC; but since it can be negative, it can accommodate situations where the NEC would fail.  By integrating the QNEC along a null generator, we can obtain the ANEC, in situations where the boundary term $S_\text{out}^{\prime}$ vanishes at early and late times.

Intriguingly, the QNEC---an intrinsically field theoretic statement---was recognized by studying conjectured properties of the generalized entropy,
\begin{equation}
S_\mathrm{gen} [\Sigma]= \frac{A[\Sigma]}{4G\hbar} + S_\text{out}[\Sigma]~,
\end{equation}
a key concept arising in quantum gravity~\cite{Bek72,Bek73,Bek74}. Here $\Sigma$ is a codimension-2 surface which divides a Cauchy surface in two, $A[\Sigma]$ is its area and $S_\text{out}$ is the von Neumann entropy of the matter fields on one side of $\Sigma$.

The generalized second law (GSL) is the conjecture~\cite{Bek72} that the generalized entropy cannot decrease as $\Sigma$ is moved up along a causal horizon. Equation~(\ref{QNEC}) first appeared as a sufficient condition for the GSL, satisfied by a nontrivial class of states of a 1+1 dimensional CFT~\cite{WallCFT}.
The QNEC emerged as a general constraint on quantum field theories when it was noted that the Quantum Focussing Conjecture (QFC) implies \eqref{QNEC} in an appropriate limit~\cite{BouFis15a}. We will briefly describe the QFC and outline how the QNEC arises from it.

A generalized entropy can be ascribed not only to horizon slices, but to any surface that splits a Cauchy surface~\cite{Wal10QST, EngWal14, BianchiMyers12, MyePou13, FLM13}. Moreover, one can define a quantum expansion $\Theta[\Sigma;y_1]$, the rate (per unit area) at which the generalized entropy changes when the infinitesimal area element of $\nu$ at a point $y_1$ is deformed in one of its future orthogonal null directions~\cite{BouFis15a} (see Fig.~\ref{fig:setup}). This quantity limits to the classical (geometric) expansion as $\hbar\to 0$. The QFC states that the quantum expansion $\Theta[\Sigma;y_1]$ will not increase under any second variation of $\Sigma$ along the same future congruence, be it at $y_1$ or at some other point $y_2$~\cite{BouFis15a}.

The QFC, in turn, was proposed as a quantum version of the covariant entropy bound (Bousso bound)~\cite{CEB1,CEB2,FMW}, a quantum gravity conjecture which bounds the entropy on a nonexpanding null surface in terms of the difference between its initial and final area. The QFC implies the Bousso bound; but because the generalized entropy appears to be insensitive to the UV cutoff~\cite{Susskind:1994sm,Jacobson:1994iw,Solodukhin:1994yz}, the QFC remains well-defined in more general settings. (The QFC is distinct from the quantum Bousso bound of \cite{BCFM1,BCFM2}, which defines the entropy by vacuum subtraction \cite{Cas08}, a procedure applicable if the gravitational effects of matter are negligible.)

In the case where $y_1 \ne y_2$, it can be shown~\cite{BouFis15a} that the QFC follows from strong subadditivity, an entropy inequality which all quantum systems must obey.\footnote{Some recent articles \cite{Bhattacharya:2014vja,Lashkari:2014kda} considered a different type of second derivative of the entropy in 1+1 field theory.  These inequalities involve varying the two endpoints of an interval independently, and therefore follow from strong subadditivity alone, without making reference to the stress-tensor.}  For $y_1=y_2$, the QFC remains a conjecture in general, but in special cases it can be proven.  The QFC constrains a combination of ``geometric'' terms proportional to $G^{-1}$ that stem from the classical expansion, as well as ``matter entropy'' terms that stem from $S_\text{out}$ and do not involve Newton's constant. The classical expansion is governed by Raychaudhuri's equation, $\theta' = -\theta^2/2 - \sigma^2 - 8\pi G \langle T_{kk} \rangle$.\footnote{Raychaudhuri's equation immediately implies that, in cases where the classical geometrical terms dominate, the QFC is true iff the classical spacetime obeys the null curvature condition.}  If the expansion $\theta$ and the shear $\sigma$ vanish at $y_1$, then the rate of change of the expansion is governed by a term proportional to $G$. In this case, all $G$'s cancel in the terms of the QFC, and \eqref{QNEC} emerges as an apparently nongravitational statement.

\paragraph{Outline}

In this paper, we will prove the QNEC in a broad arena. Our proof applies to free or superrenormalizable, massive or massless bosonic fields, in all cases where the surface $\Sigma$ lies on a stationary null hypersurface (one with everywhere vanishing expansion). The most important example is Minkowski space, with $\Sigma$ lying on a Rindler horizon. Such a horizon exists at every point $p$, with every orientation $k^a$, so the QNEC constrains all null components of the stress tensor everywhere in Minkowski space.

A similar situation arises in a de~Sitter background, where $p$ and $k^a$ specify a de~Sitter horizon, and in Anti-de~Sitter space, where they specify a Poincar\'e horizon.  Other examples include an eternal Schwarzschild or Kerr black hole, but in this case our proof applies only to points on the horizon, with $k^a$ tangent to the horizon generators. These should all be viewed as fixed background spacetimes with no dynamical gravity; our proof establishes that free scalar field theory on these backgrounds satisfies \eqref{QNEC}.

We give a brief review of the formal statement of the QNEC in Sec.~\ref{sec-review}. We then set up the calculation of all relevant terms in Sec.~\ref{sec-reducetocft}. In Sec.~\ref{sec-nullquant}, we review the null surface quantization of the theory, on the particular null surface $N$ that is orthogonal to $\Sigma$ with tangent vector $k^a$.  Null quantization has the remarkable feature that the vacuum state factorizes in the transverse spatial directions. This reduces any purely kinematic problem (such as ours) to the analysis of a large number of copies of the free chiral scalar CFT in 1+1 dimensions. We then restrict attention to the particular chiral CFT on the infinitesimal pencil that passes through the point $p$ where $\Sigma$ is varied.  The state on this pencil is entangled with an auxiliary quantum system which contains both the information crossing the other generators of $N$, and the information that does not fall across $N$ at all.

In the 1+1 chiral CFT, the pencil state is very close to the vacuum, but not so close that the QNEC would be trivially saturated by application of the first law of the entanglement entropy. To constrain the second order variations of $S_\text{out}$ (the Fisher information), we must keep track of the deviation of the pencil state from the vacuum to second order. We discuss the appropriate expansion of the overall state in Sec.~\ref{sec-expandrho}. We write the state in terms of operators inserted on the Euclidean plane corresponding to the pencil and expand in a basis of the auxiliary system. Then in Sec.~\ref{sec-expandS}, we expand the entropy and identify the parts of our expnsion enter into the second derivative.

In Sec.~\ref{sec-3}, we compute the sign of $\langle T_{kk} \rangle - \frac{\hbar}{2\pi {\cal A}} S_\text{out}^{ \prime\prime} $. In Sec.~\ref{sec:generalReplicaTrick} we review the replica trick for computing the von Neumann entropy by the analytic continuation of Renyi entropies. We extract two terms relevant to the QNEC, which are computed in Sec.~\ref{sec-t1} and~\ref{sec-t2} respectively. The most subtle part of the calculation is the analytic continuation of the second of these terms, in Sec.~\ref{app:analyticCont}. In Sec.~\ref{putItTogether}, we combine the terms and conclude that the QNEC holds for all states.

In Sec.~\ref{sec-extensions}, we extend our result to establish the QNEC also for superrenormalizable scalar fields, and for bosonic fields of higher spin. We also discuss the extension to interacting theories.  We expect that the proof we have given can be extended to fermionic fields, but we leave this task for the future.

\paragraph{Discussion}

Our result establishes a new and surprising link between quantum information and a more familiar physical quantity, the stress tensor. The QNEC identifies the ``acceleration'' of information transfer as a lower bound on the energy density. Equivalently, the stress tensor can be viewed as imposing a constraint on the second derivative of the von Neumann entropy. The latter can be difficult to calculate but plays an important role in quantum information theory, condensed matter, and high energy physics.

Our proof of the QNEC requires no assumptions beyond the known properties of free quantum fields, but it is quite lengthy and somewhat involved. Yet, the QNEC follows almost trivially from a statement involving gravity, the Quantum Focussing Conjecture. This perplexing situation is somewhat reminiscent of the proof of the quantum Bousso bound~\cite{BCFM1}, particularly in the interacting case~\cite{BCFM2}. It is intriguing that the study of quantum gravity can lead us to simple conjectures such as \eqref{QNEC} which can be proven entirely within the nongravitational sector, where they are far from obvious---so far, indeed, that they had not been recognized until they emerged as implications of holographic entropy bounds or of properties of the generalized entropy.

It is becoming clear that the structure of known quantum field theories carries a deep imprint of causal and information theoretic properties ultimately dictated by quantum gravity. This adds to the evidence that ``quantizing gravity'' has nothing to do with the inclusion of one last force in a quantization program. It would be interesting to try to formulate models of quantum gravity in which focussing of the entropy occurs naturally.



Remarkably, the QNEC does not seem to follow from any of the standard identities that apply purely at the level of quantum information. Our proof did involve additional structure supplied by quantum field theory.  The QNEC is related to the relative entropy $S(\rho|\sigma) = \trace(\rho \ln \rho) - \trace(\rho \ln \sigma)$, which equals $-S_\text{gen}$ (up to a constant) when $\sigma$ is taken to be the vacuum state. The relative entropy satisfies positivity, which guarantees that $S_\mathrm{gen}(\rho)$ is less than in the vacuum state. It also enjoys monotonicity, which implies that $S_\mathrm{gen}$ is increasing under restrictions; this constrains the first derivative, which is the GSL \cite{Wall11}.  It may appear that the QNEC can be proven using properties of the relative entropy.  But the QNEC is a statement about the \emph{second derivative} of the generalized entropy.  It is possible that the QNEC hints at more general quantum information inequalities, which are yet to be discovered.  It is interesting that a recently proposed new GSL, which applies in strongly gravitating regions such as cosmology, also can be shown to follow from the QFC~\cite{BouEngTA}.

\section{Statement of the Quantum Null Energy Condition}
\label{sec-review}

The statement of the QNEC involves the choice of a point $p$ a null vector $k^a$ at $p$, and a smooth codimension-2 surface $\Sigma$ orthogonal to $k^a$ at $p$ such that $\Sigma$ splits a Cauchy surface into two portions. The null vector $k^a$ is a member of a vector field orthogonal to $\Sigma$ defined in a neighborhood of $p$, $k^a(y)$. Here and below we use $y$ as a coordinate label on $\Sigma$, also called the ``transverse direction." We can consider a family of surfaces $\Sigma[\lambda(y)]$ obtained by deforming $\Sigma$ along the null geodesics generated by $k^a(y)$ by the affine parameters $\lambda(y)$. 


The deformed surfaces will also be Cauchy-splitting \cite{BouEng15b}. This allows us to define a family of entropies $S_{\rm out}[\lambda(y)]$, which are the von Neumann entropies of the quantum fields restricted to the Cauchy surface on one side of $\Sigma[\lambda(y)]$. The choice of Cauchy surface is unimportant, since by unitarity the entropy will be independent of that choice. The choice of side of $\Sigma[\lambda(y)]$ also does not matter, because the QNEC is symmetric with respect to \(k^{a} \to -k^{a}\). 

Once we have defined $S_\text{out}[\lambda(y)]$, we can consider its functional derivatives. In general, the second functional derivative will contain diagonal and off-diagonal terms (present because $S_{\rm out}$ is a non-local functional), and the diagonal terms will be proportional to a $\delta$-function. We define the second functional derivative at coincident points by factoring out that $\delta$-function:
\be
\frac{\delta^2 S_{\rm out}}{\delta \lambda(y)\delta \lambda(y')} = \frac{\delta^2 S_{\rm out}}{\delta\lambda(y)^2}\delta(y-y') + \text{off-diagonal}.
\ee
Then if the expansion and the shear of $k^a(y)$ vanish at $p$, we have the general conjecture
\be\label{eq-generalQNEC}
\braket{T_{kk}(p)} \geq \frac{\hbar}{2\pi \sqrt{h(p)}} \frac{\delta^2 S_{\rm out}}{\delta\lambda(p)^2}\Big|_{\lambda(y) =0}~,
\ee
where $h$ is the determinant of the induced metric on $\Sigma$ and $T_{kk} \equiv T_{ab}k^ak^b$. We will find it convenient below to work with a discretized version of the functional derivative, obtained by dividing $\Sigma$ into regions of small area $\cal A$ and considering variations locally constant in those regions. Then \eqref{eq-generalQNEC} reduces to the form advertised in \eqref{QNEC}:
\begin{equation}
\langle T_{kk} \rangle \ge \frac{\hbar}{2\pi {\cal A}} S_\text{out}^{ \prime\prime}~.
\end{equation}

\section{Reduction to a 1+1 CFT and Auxiliary System}
\label{sec-reducetocft}

\subsection{Null Quantization}
\label{sec-nullquant}

The proof that follows applies when $\Sigma$ is a section of a general stationary null surface $N$ in $D > 2$ (the case $D = 2$ will be treated separately, in section~\ref{sec-extensions}).
We consider deformations of $\Sigma$ along $N$ toward the future, so the deformation vector $k^a$ is future-directed, and we choose to take the ``outside'' direction to be the side towards which $k^a$ points. As mentioned above, a proof of this case automatically implies a proof for the opposite choice of outside.
By unitary time evolution of the spacelike Cauchy data, we can consider the state to be defined on the portion of $N$ in the future of $\Sigma$ together with a portion of future null infinity. 

We rely on null quantization on $N$, which requires that $N$ be stationary~\cite{Wall11}. 
Null quantization is simplest if we first discretize $N$ along the transverse direction into regions of small transverse area $\calA$. These regions, which are fully extended in the null direction, are called pencils. Ultimately we will take the continuum limit $\calA \to 0$, and the QNEC will be shown to hold in this limit. At intermediate stages, $\calA$ acts as a small expansion parameter.\footnote{The dimensionless expansion parameter is $\calA$ in units of a characteristic length scale of the state we are interested in, e.g., the wavelength of typical excitations. The state remains fixed as $\calA\to 0$.} This is the reason why we are restricting ourselves to $D>2$ spacetime dimensions for now: without a transverse direction to discretize, there would be no small expansion parameter. Also, while logically independent from the discretization used to define the QNEC in \eqref{QNEC}, we will take these two discretizations to be the same. That is, we will consider deformations of the surface $\Sigma$ which are localized to the same regions of size $\calA$ that define the discretized null quantization.

There is a distinguished pencil that contains the point $p$; this is the pencil on which we will perform our deformations. The total Hilbert space of the system can be decomposed as $\mathcal{H} = \mathcal{H}_{\rm pen} \otimes \mathcal{H}_{\rm aux}$, where $\mathcal{H}_{\rm pen}$ refers to the fields on the distinguished pencil and $\mathcal{H}_{\rm aux}$ is everything else. ``Everything else'' includes both the remaining pencils on $N$ restricted to the future of $\Sigma$, as well as the relevant portion of null infinity. We do not have to be specific about the exact structure of the auxiliary system; our proof does not assume anything about it other than what is implied by quantum mechanics. Beginning with a density matrix on $\mathcal{H}$, we obtain a one-parameter family of density matrices $\rho(\lambda)$ by tracing out the part of the pencil in the past of affine parameter $\lambda$. When $\lambda\to -\infty$ the pencil is fully extended, and when $\lambda \to +\infty$ the entire pencil has been traced out. $\lambda=0$ corresponds to no deformation of the original surface.

When restricted to $N$, the theory decomposes into a product of 1+1-dimensional free chiral CFTs, with one CFT associated to each pencil of $N$. In particular, this means that the vacuum state factorizes with respect to the pencil decomposition of $N$ \cite{Wall11}. 

Crucially, when $\calA$ is small, the state of the pencil is near the vacuum. This can be seen as follows. For a region of small size $\calA$, the amplitude to have $n$ particles on the pencil scales like $\calA^{n/2}$ (so the probability is appropriately extensive), and therefore the coefficient of $|n\rangle\!\langle m|$ in the pencil Fock basis expansion of the state scales like $\calA^{(n+m)/2}$. Hence for small $\calA$ we can write the state as
\begin{equation}\label{eq-state}
\rho(\lambda) = \rho^{(0)}_{\rm pen}(\lambda) \otimes \rho^{(0)}_{\rm aux} + \sigma(\lambda)~,
\end{equation}
where $\rho^{(0)}_{\rm pen}(\lambda)$ is the vacuum state density matrix on the part of the pencil with affine parameter greater than $\lambda$, $\rho^{(0)}_{\rm aux}$ is some state in the auxiliary system (not necessarily the vacuum), and the perturbation $\sigma(\lambda)$ is small: the largest terms are obtained by taking the partial trace of $|0\rangle\!\langle 1|$ and $|1\rangle\!\langle 0|$ in the pencil Fock basis, and these terms have coefficients which scale like $\calA^{1/2}$. Entanglement between the pencil and the auxiliary system is also present in $\sigma$; we will explore the form of $\sigma$ in more detail in the following section.
 
\subsection{Expansion of the State}
\label{sec-expandrho}

As discussed above, the pencil state can be described in terms of a 1+1-dimensional free chiral CFT, with fields that depend only on the coordinate $z= x+t$. In this notation, translations along the Rindler horizon in the 1+1 CFT are translations in \(z\), and are generated by \(\partial \equiv \frac{\partial}{\partial z}\). In a chiral theory, this is equivalent to translations in the spatial coordinate \(x\). Therefore the shift in affine parameter $\lambda$ of the previous section can be replaced by a shift in the spatial coordinate for the purposes of the CFT calculation. In addition, quantization on a surface of constant Euclidean time \(\tau= it = 0\) in a chiral theory is equivalent to quantization on the Rindler horizon. Thus when we construct the state we can use standard Euclidean methods for two-dimensional CFTs.

\begin{figure}[t]
	\centering
	\includegraphics{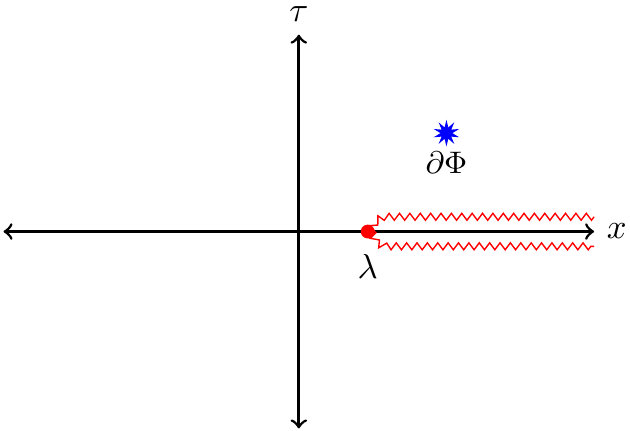}
	\caption{The state of the CFT on $x > \lambda$ can be defined by insertions of $\partial \Phi$ on the Euclidean plane. The red lines denote a branch cut where the state is defined.}
	\label{fig:xtauplane}
\end{figure}

We have argued that, at order $\calA^{1/2}$, the perturbation $\sigma$ on the full pencil must be of the schematic form $\oprod{0}{1}$ (plus Hermitian conjugate). So on the full pencil, we have the state
\be\label{eq-fullpencilstate}
\rho = \rho(-\infty) = \oprod{0}{0} \otimes \rho^{(0)}_{\rm aux} +\calA^{1/2} \sum_{ij} \left(\oprod{0}{\psi_{ij}}+\oprod{\psi_{ji}}{0}\right) \otimes \oprod{i}{j} + \cdots,
\ee
where $\oprod{i}{j}$ is a basis of operators in the auxiliary system and ``$\cdots$" denotes terms which vanish more quickly as $\calA\to0$. We will argue in Sec.~\ref{sec-expandS} that those terms are not relevant for the QNEC, and so we will ignore them from now on. For later convenience, we will take the basis $\ket{i}$ in the auxiliary system to be the one in which $\rho^{(0)}_{\rm aux}$ is diagonal. The states $\ket{\psi_{ij}}$ are single-particle states in the CFT, and we have ensured that the state is Hermitian. The CFT part of the state can be constructed by acting on the vacuum with a single copy of the field operator. In a Euclidean path integral picture, we can get the most general single-particle state by allowing arbitrary single-field insertions on the Euclidean plane. This is shown in Fig.~\ref{fig:xtauplane}.

To obtain the state at a finite value of $\lambda$, we need to take the trace of \eqref{eq-fullpencilstate} over the region $x<\lambda$. Alternatively, we can hold fixed the inaccessible region, $x<0$, but translate the field operators used to construct the state by $\lambda$. From this point of view the vacuum is independent of $\lambda$ and we write it as
\be
	\rho^{(0)}_{\rm pen} = e^{-2\pi K_{\rm pen}},
\ee
where, up to an additive constant, the modular Hamiltonian $K_{\rm pen}$ coincides with the Rindler boost generator for the CFT~\cite{Unr76, BisWic76}. Specializing to the case of a single chiral scalar field (extensions will be discussed in Sec.~\ref{sec-extensions}), the trace of \eqref{eq-fullpencilstate} becomes
\begin{equation}\label{eq-tempstate}
	\rho(\lambda) = e^{-2\pi K_{\rm pen}} \otimes \rho^{(0)}_{\rm aux} +\calA^{1/2} \sum_{ij} \left(e^{-2\pi K_{\rm pen}}\int dr d\theta\ f_{ij}(r,\theta) \partial\Phi(re^{i\theta} - \lambda)\right) \otimes \oprod{i}{j},
\end{equation}
where $\partial\Phi(z)$ is now a holomorphic local operator on a two-dimensional Euclidean plane\footnote{We insert \(\partial\Phi\) instead of \(\Phi\) in order to remove any zero-mode subtleties. We have checked that the proof still works formally if one inserts $\Phi$ instead of $\partial \Phi$, and in fact continues to work when an arbitrary number of derivatives, $\partial^l\Phi$, are used. This latter fact is not surprising since insertions of $\Phi$ alone (or $\partial \Phi$ if we drop the zero mode) are sufficient to generate all single particle states. See \cite{Burkardt:1995ct, Wall11} for details on the zero-mode.} and $(r,\theta)$ are polar coordinates on that plane, with $z=re^{i\theta}$. Rotations in $\theta$ are generated by $K_{\rm pen}$. Thus the operator \(\partial\Phi\) is defined by\footnote{Here $\theta$ is restricted to be in the range $[0,2\pi)$.}
\begin{align}\label{phases}
	\partial\Phi(re^{i\theta}) = e^{-i\theta} e^{\theta K_{\text{pen}}} \partial \Phi(r) e^{-\theta K_{\text{pen}}}.
\end{align}
All of the operators in \eqref{eq-tempstate} are manifestly operators on the Hilbert space corresponding to \(x>0, \tau=0\). We are taking $\Phi$ to be a real scalar field, so in particular $\partial \Phi$ is a Hermitian operator for real arguments. Then in order for $\rho$ to be Hermitian, we must have
\be\label{hermThetaConstraint}
	f_{ij}(r,\theta) = f_{ji}(r,2\pi -\theta)^*.
\ee
Aside from this reality condition, letting $f$ be completely general gives all possible single particle states.

To facilitate our later calculations, we will modify \eqref{eq-tempstate} in order to put the auxiliary system on equal footing with the CFT. To that end, define $K_{\rm aux}$ through the equation $\rho^{(0)}_{\rm aux} = \exp(-2\pi K_{\rm aux})$. We can invent a coordinate $\theta$ for the auxiliary system and declare that evolution in $\theta$ is generated by $K_{\rm aux}$. Then define the operators
\be
E_{ij}(\theta) \equiv e^{\theta K_{\rm aux}}\oprod{i}{j}e^{-\theta K_{\rm aux}} = e^{\theta (K_i-K_j)}\oprod{i}{j}. 
\ee
Since $K_{\rm aux}$ is diagonal in the $\ket{i}$ basis, with eigenvalues $K_i$, $E_{ij}(\theta)$ is just a rescaled $\oprod{i}{j}$. More generally, multiplying $\oprod{i}{j}$ on either side by arbitrary functions of $K_{\rm aux}$ results in the same operator up to an $(i,j)$-dependent numerical factor. So by making the replacement
\be
f_{ij}(r,\theta) \to e^{(2\pi-\theta) K_i}e^{\theta K_j} f_{ij}(r,\theta)~,
\ee
which does not alter the reality condition on $f$, we can write
\begin{equation}\label{eq-state2}
\rho(\lambda) = e^{-2\pi K_{\rm tot}} +\calA^{1/2}  e^{-2\pi K_{\rm tot}}\sum_{ij} \int \! dr \, d\theta\, f_{ij}(r,\theta) \partial\Phi(re^{i\theta} - \lambda) \otimes E_{ij}(\theta)~,
\end{equation}
where $K_{\rm tot} \equiv K_{\rm pen} + K_{\rm aux}$. From now on, we will simply write $K$ for $K_{\rm tot}$.

Below it will be useful to write \(\sigma(\lambda)\) as
\begin{align}\label{eq-Odef}
	\sigma(\lambda) \equiv \mathcal{A}^{1/2} \rho^{(0)} \mathcal{O}(\lambda)~.
\end{align}
Thus comparing with \eqref{eq-state2}, we find
\begin{align}\label{eq-Oexpr}
	\mathcal{O(\lambda)} = \sum_{ij} \int \! dr \,d\theta\, f_{ij}(r,\theta) \partial\Phi(re^{i\theta} - \lambda) \otimes E_{ij}(\theta)~.
\end{align}

As a side comment, we note that one could prepare the state \eqref{eq-state2} via a Euclidean path integral over the entire plane with an insertion of \(\mathcal{O}\) and boundary field configurations defined at \(\theta = 0^{+}\) and \(\theta = (2\pi)^{-}\).


\subsection{Expansion of the Entropy}
\label{sec-expandS}

In the previous sections we saw that null quantization gives us a state of the form
\begin{equation}
\rho(\lambda) = \rho^{(0)}_{\rm pen}(\lambda) \otimes \rho^{(0)}_{\rm aux} + \sigma(\lambda),
\end{equation}
where $\rho^{(0)}_{\rm pen}(\lambda)$ is the vacuum state reduced density matrix on the part of the pencil with affine parameter greater than $\lambda$, $\rho^{(0)}_{\rm aux}$ is an arbitrary state in the auxiliary system, and the perturbation $\sigma$ is proportional to the small parameter $\calA^{1/2}$. In this section, we will expand the entropy perturbatively in $\sigma$ and show that the QNEC reduces to a statement about the contributions of $\sigma$ to the entropy. We will assume that both $\rho(\lambda)$ and $\rho^{(0)}(\lambda) \equiv \rho^{(0)}_{\rm pen}(\lambda) \otimes \rho^{(0)}_{\rm aux}$ are properly normalized density matrices, so ${\trace(\sigma) = 0}$.

The von Neumann entropy of $\rho(\lambda)$ is $S_\text{out}(\lambda)$. We will expand it as a perturbation series in $\sigma(\lambda)$:
\be\label{eq-Slambda}
S_{\rm out}(\lambda) = S^{(0)}(\lambda) + S^{(1)}(\lambda) + S^{(2)}(\lambda) + \cdots
\ee
where $S^{(n)}(\lambda)$ contains $n$ powers of $\sigma(\lambda)$. At zeroth order, since $\rho^{(0)}$ is a product state, we have
\begin{equation}
S^{(0)}(\lambda) = - \trace\left[\rho^{(0)}(\lambda) \log \rho^{(0)}(\lambda)\right] =  - \trace\left[\rho_{\rm pen}^{(0)}(\lambda) \log \rho_{\rm pen}^{(0)}(\lambda)\right]- \trace\left[\rho_{\rm aux}^{(0)} \log \rho_{\rm aux}^{(0)}\right].
\end{equation}
The first term on the right-hand side is independent of $\lambda$ because of null translation invariance of the vacuum: all half-pencils have the same vacuum entropy. The second term is manifestly independent of $\lambda$. So $S^{(0)}$ is $\lambda$-independent and does not play a role in the QNEC.

Now we turn to $S^{(1)}(\lambda)$:
\begin{equation}
S^{(1)}(\lambda) = -\trace\left[\sigma(\lambda) \log \rho^{(0)}(\lambda)\right] = -\trace\left[\sigma(\lambda) \log \rho_{\rm pen}^{(0)}(\lambda)\right] -\trace\left[\sigma(\lambda) \log \rho_{\rm aux}^{(0)}\right].
\end{equation}
Once again, the second term is $\lambda$-independent, which we can see by evaluating the trace over the pencil subsystem:
\be
\trace\left[\sigma(\lambda) \log \rho_{\rm aux}^{(0)}\right] = \trace_{\rm aux}\left[\left[\trace_{\rm pen}\sigma(\lambda) \right]\log \rho_{\rm aux}^{(0)}\right] = \trace_{\rm aux}\left[\sigma(\infty)\log \rho_{\rm aux}^{(0)}\right].
\ee
To evaluate the first term, we use the fact that $\rho^{(0)}_{\rm pen}(\lambda)$ is thermal with respect to the boost operator on the pencil. Then we have
\be
-\trace\left[\sigma(\lambda) \log \rho^{(0)}_{\rm pen}(\lambda)\right] = \frac{2\pi \calA}{\hbar} \int_\lambda^\infty d\lambda'  \,(\lambda'-\lambda)\langle T_{kk}(\lambda')\rangle,
\ee
where the integral is along the generator which defines the pencil and the expectation value is taken in the excited state. This is the first $\lambda$-dependent term we have in the perturbative expansion of $S(\lambda)$. Taking two derivatives and evaluating at $\lambda=0$ gives the identity
\be
\left(S^{(0)} + S^{(1)} \right)'' =  \frac{2\pi \calA}{\hbar}\langle T_{kk}\rangle.
\ee
Subtracting $S_{\rm out}''$ from both sides of this equation shows that
\be\label{QNECandS2}
\frac{\hbar}{2\pi \calA}S_{\rm out}''-\langle T_{kk}\rangle = \frac{\hbar}{2\pi \calA} \left(S_{\rm out}-S^{(0)} - S^{(1)}\right)'' = \frac{\hbar}{2\pi \calA}{S^{(2)\dprime}}+ \cdots,
\ee
where ``$\cdots$" contains terms higher than quadratic order in $\sigma$. The QNEC (equation~\eqref{QNEC}) is the statement that this quantity is negative in the limit $\calA\to 0$. Earlier we showed that $\sigma$ was proportional to $\calA^{1/2}$. Then $S^{(2)}$ is proportional to $\calA$, and we must check that ${S^{(2)\dprime}}$ is negative. However, the higher order terms $S^{(\ell)}$ for $\ell>2$ vanish more quickly with $\calA$ and therefore drop out in the limit $\calA\to 0$.

We have shown that the QNEC reduces to the statement that ${S^{(2)\dprime}}\leq 0$ for perturbations from the vacuum. In fact, we have shown something a little stronger. In general, the perturbation $\sigma$ will have terms proportional to $\calA^{n/2}$ for all $n\geq1$. Our arguments show that only the term proportional to $\calA^{1/2}$ matters for the QNEC, and furthermore that this term is off-diagonal in the single-particle/vacuum subspace. So we can simplify matters by considering states which contain {\em only} such a term proportional to $\calA^{1/2}$ and no higher powers of $\calA$. In other words, we can take the state to be of the form in \eqref{eq-fullpencilstate} with the unwritten ``$\cdots$" terms set equal to zero. Now we only need to show that ${S^{(2)\dprime}}\leq 0$ for such states.

%
%
%

\section{Calculation of the Entropy}
\label{sec-3}

\subsection{The Replica Trick} \label{sec:generalReplicaTrick}
The replica trick prescription is to use the following formula for the von Neumann entropy~\cite{Callan:1994py}:
\begin{align}
	S_{\text{out}} = - \trace[\rho \log{\rho}] = (1-n\partial_{n}) \log{\trace[\rho^{n}]} \Big|_{n=1}~.
\end{align}
This can be written as
\begin{align}\label{replicaTrick}
	S_{\text{out}} = \replicaoperator \log{\tilde Z_{n}}
\end{align}
where \(\tilde Z_{n} \equiv \trace[\rho^{n}]\)\footnote{In the replica trick one often works with the partition function $Z_n$, in terms of which $\tilde Z_n = Z_n/(Z_1)^n$. Choosing $Z_n$ over $\tilde Z_n$ is equivalent to choosing a different normalization for $\rho$, but we find it convenient to keep $\trace \rho =1$.} and the operator \(\replicaoperator\) is defined by
\begin{align}
	\replicaoperator f({n}) \equiv (1-n\partial_{n}) f({n}) \big|_{n=1}
\end{align}
where \(f({n})\) is some function of \(n\). Since $\tilde Z_n$ is only defined for integer values of $n$, we first must analytically continue to real $n>0$ in order to apply the $\replicaoperator$ operator. The analytic continuation step is in general quite tricky, and will require care in our calculation. (Our analytic continuation is performed in Section~\ref{app:analyticCont}.) 

On general grounds discussed above, we must study the second-order term in a perturbative expansion of the entropy about the state $\rho^{(0)}$. Suppressing all \(\lambda\) dependence, we have
\be
\tilde Z_n = \trace \left[ (\rho^{(0)} + \sigma)^n\right]~.
\ee
Expanding $\tilde Z_n$ to quadratic order to isolate \({S^{(2)}}''\), we have
\be
\tilde Z_n = \trace \left[(\rho^{(0)})^n\right]+ n\trace \left[  \sigma (\rho^{(0)})^{n-1} \right]+ \frac{n}{2}\sum_{k=0}^{n-2}\trace\left[ (\rho^{(0)})^k\sigma(\rho^{(0)})^{n-k-2}\sigma\right] + \cdots.
\ee
Using the notation introduced in \eqref{eq-Odef} we can write
\be\label{eq-traceexpand}
\tilde Z_n = \trace \left[(\rho^{(0)})^n\right]+ n\trace \left[  \mathcal{O}(\rho^{(0)})^n \right]+ \frac{n}{2}\sum_{k=1}^{n-1}\trace\left[ (\rho^{(0)})^{-k}\mathcal{O}(\rho^{(0)})^{k}\mathcal{O}(\rho^{(0)})^{n}\right] + \cdots.
\ee
We denote by $\mathcal{O}^{(k)}$ the operator $\mathcal{O}$ conjugated by $(\rho^{(0)})^k$:
\begin{align}\label{eq-Okdef}
	\mathcal{O}^{(k)} &\equiv (\rho^{(0)})^{-k}\mathcal{O}(\rho^{(0)})^{k}\\
	&= e^{2\pi k K} \mathcal{O} e^{-2\pi k K}~.
\end{align}
This is equivalent to a Heisenberg evolution of $\mathcal{O}$ in the angle $\theta$ by an amount $2\pi k$. Since $\mathcal{O}$ is the integral of operators with angles $0\leq \theta < 2\pi$, it follows that $\mathcal{O}^{(k)}$ will be an integral over operators with angles $2\pi k < \theta < 2\pi (k+1)$.\footnote{One could worry that the phase factor in \eqref{phases} spoils this relation, but notice that the phase has period $2\pi$ in $\theta$ and so does not appear when shifting by $2\pi k$.} Furthermore, since rotations by $2\pi k$ commute with translations by $\lambda$, we can obtain $\mathcal{O}^{(k)}$ from $\mathcal{O}$ simply by letting the range of integration that defines $\mathcal{O}$ shift from $[0,2\pi]$ to $[2\pi k, 2\pi (k+1)]$, as long as we define $f_{ij}(r,\theta)$ to be periodic in $\theta$ with period $2\pi$.

It will also be convenient to introduce an angle-ordered expectation value, defined as
\be
\braket{\dots}_{n} \equiv \frac{\trace[(\rho^{(0)})^{n} \thetaorder{\dots}]}{\trace[(\rho^{(0)})^{n}]},
\ee
where $\thetaorder{\dots}$ is $\theta$-ordering. Then \eqref{eq-traceexpand} can be written
\be
\tilde Z_n = \trace \left[(\rho^{(0)})^n\right] \left(1+ n\ev{ \mathcal{O}}_n + \frac{n}{2} \sum_{k=1}^{n-1} \ev{\mathcal{O}^{(k)}\mathcal{O}}_n\right) + \cdots.
\ee
Taking the logarithm of $\tilde Z_n$ and extracting the part quadratic in $\sigma$ gives
\be\label{eq-intermediateZ}
\log \tilde Z_n \supset  \frac{n}{2} \sum_{k=1}^{n-1} \ev{\mathcal{O}^{(k)}\mathcal{O}}_n - \frac{n^2}{2}\ev{ \mathcal{O}}_n^2,
\ee
where we have kept only the part quadratic in $\mathcal{O}$. The contribution of the second term to the entanglement entropy will be proportional to $\ev{\mathcal{O}}$, which vanishes because of the tracelessness of $\sigma$. Therefore we only need to consider the first term.

Since we are considering angle-ordered expectation values, we have the identity
\be
\ev{{\left(\sum_{k=0}^{n-1}\mathcal{O}^{(k)}\right)^2}}_n  =  n\sum_{k=0}^{n-1}\ev{{\mathcal{O}^{(k)}\mathcal{O}}}_n,
\ee
and so from the first term in \eqref{eq-intermediateZ} the relevant part of $\log \tilde Z_n$ can be written as
\be\label{discreteO2ptFunction-1}
\log \tilde Z_n \supset - \frac{n}{2}\ev{{\mathcal{O}\mathcal{O}}}_n + \frac{1}{2}\ev{{\left(\sum_{k=0}^{n-1}\mathcal{O}^{(k)}\right)^2}}_n .
\ee
Restoring the \(\lambda\) dependence and taking \(\lambda\) derivatives gives
\begin{align}
	{S^{(2)\dprime}}&= \frac{\partial^{2}}{\partial \lambda^{2}} \bigg|_{\lambda=0} \replicaoperator \log{\tilde Z_{n}(\lambda)}\\
	&= \replicaoperator\ \frac{-n}{2}\ev{{\mathcal{O}\mathcal{O}}}''_n + \replicaoperator\ \frac{1}{2}\ev{{\left(\sum_{k=0}^{n-1}\mathcal{O}^{(k)}\right)^2}}''_n. \label{discreteO2ptFunction}
\end{align}
The \(\braket{\dots}_{n}''\) notation means take two \(\lambda\) derivatives and then set \(\lambda=0\). In the following sections we will compute these two terms separately.

We note that the two terms in \eqref{discreteO2ptFunction} are analogous to $\delta S_{EE}^{(1)}$ and $\delta S_{EE}^{(2)}$ of Ref.~\cite{Fau14}, where a similar perturbative computation of the entropy was performed. Though the details of the two calculations differ (in particular we have an auxiliary system as well as a CFT), it would be interesting to explore further the connection between our present work and that of Ref.~\cite{Fau14}.

\subsection{Evaluation of Same-Sheet Correlator}
\label{sec-t1}
In this section we consider the term \(\braket{{\mathcal{O}\mathcal{O}}}_{n}''\) appearing in \eqref{discreteO2ptFunction}. The analytic continuation of this term in \(n\) is straightforward. We first apply \(\replicaoperator\):
\begin{align}
	\replicaoperator \frac{-n}{2} \braket{{\mathcal{O}\mathcal{O}}}_{n} &= \replicaoperator\ \frac{-n}{2} \frac{\trace \left[ e^{-2\pi nK} \thetaorder{\mathcal{O}\mathcal{O} }\right]}{\trace [e^{-2\pi n K}]} \\
	&= - \pi \braket{{ \mathcal{O}\mathcal{O}}\Delta K}\label{DfirstTerm}
\end{align}
where $\Delta K \equiv K - \braket{K}$ is the vacuum-subtracted modular Hamiltonian. When an expectation value \(\braket{\dots}\) appears without a subscript it is understood to refer to the normalized expectation value \(\braket{\dots}_{n}\) with \(n=1\), i.e., the angle-ordered expectation value with respect to $\rho^{(0)}$. Also note that $K$ appears outside of the angle-ordering in the trace form of the expectation value, which is formally equivalent to being inserted at $\theta=0$.

We now consider the \(\lambda\) dependence. Recall that $K$ is defined to be $\lambda$-independent, and the $\lambda$-dependence of $\mathcal{O}$ enters through a shift in the coordinate insertion of $\partial \Phi$ (see \eqref{eq-Oexpr}). We first split $\Delta K$ into $\Delta K_{\rm pen}$ and $\Delta K_{\rm aux}$. The expectation value involving $\Delta K_{\rm aux}$ will be independent of $\lambda$ because of translation invariance of the CFT, and so can be ignored. Since $K_{\rm pen}$ is the CFT boost generator on the half-line $x>0$, $\Delta K_{\rm pen}$ has a well-known expression in terms of the energy-momentum tensor of the CFT~\cite{Unr76, BisWic76}:
\be
\Delta K_{\rm pen} = \calA\int_0^\infty dx  \, x\, T_{kk}(x)= -\frac{1}{2\pi}\int_{0}^{\infty} dx\, x \, T(x)~.
\ee
Therefore the correlation function \eqref{DfirstTerm} is expressed in terms of the correlation functions $\ev{\partial\Phi(z-\lambda)\partial\Phi(w-\lambda) T(x)}$, which are the same as $\ev{\partial\Phi(z)\partial\Phi(w)T(x + \lambda)}$ by translation invariance. This makes the $\lambda$-derivatives easy to evaluate. We find
\begin{align}\label{eq:ooprimeprime}
	\replicaoperator \frac{-n}{2}\braket{{\mathcal{O}\mathcal{O}}}_{n}'' = \frac{1}{2} \braket{{\mathcal{O}\mathcal{O}}T(0)}.
\end{align}
Inserting the explicit form of \(\mathcal{O}\) gives
\begin{align}\label{Texpr2}
	\braket{ {\mathcal{O} \mathcal{O}}T(0)}
	= \frac{1}{(2\pi)^{2}} \sum_{\substack{i,j,i'j' \\ m,m'}} \int \! dr \, dr' \, d\theta \, d\theta'\ \Big( &f^{(m)}_{ij}(r) f^{(m')}_{i'j'}(r') e^{-im\theta} e^{-im'\theta'}  \nonumber\\[-1em]
	& \times \braket{\partial\Phi(re^{i\theta})\partial\Phi(r'e^{i\theta'})T(0) }\braket{{E_{ij}(\theta)E_{i'j'}(\theta')}}\Big)~,
\end{align}
where we have introduced Fourier representations of \(f_{ij}(r,\theta)\) defined by
\be\label{fourierf'sn=1}
f_{ij}(r,\theta) = \frac{1}{2\pi} \sum_{m=-\infty}^\infty f_{ij}^{(m)}(r) e^{-i m \theta}~.
\ee
 The correlation functions we need are evaluated in the appendix. Plugging equation \eqref{thermalExpect} with \(n=1\) and equation \eqref{Tphiphi} into equation \eqref{Texpr2} yields
\begin{align}
	& \braket{{\mathcal{O}\mathcal{O}}T(0)}\nonumber \\
	&= \frac{-2}{(2\pi)^{3}} \sum_{\substack{i,j,p\\ m,m'}} \int \frac{dr \, dr' \, d\theta \, d\theta' }{(rr')^{2}} f^{(m)}_{ij}(r) f^{(m')}_{ji}(r') e^{-\pi(K_{i}+K_{j})}\frac{\sinh{\pi\alpha_{ij}}}{ip+\alpha_{ij}}e^{i\theta(-p-m-2)}e^{i\theta'(p-m'-2)}\nonumber \\
	&=\frac{1}{\pi} \sum_{i,j,m} \int \frac{dr \, dr'}{(rr')^{2}} f^{(m-2)}_{ij}(r) f^{(-m-2)}_{ji}(r') e^{-\pi(K_{i}+K_{j})}\frac{\sinh{\pi\alpha_{ij}}}{im-\alpha_{ij}}, \label{finalTResult}
\end{align}
where we used the Kronecker deltas coming from the \(\theta\) integration and redefined the dummy variable \(m \to m-2\), and $\alpha_{ij} \equiv K_i - K_j$ is the difference between two eigenvalues of $K_{\rm aux}$. Note that we reserve the letters $p$ and $q$ throughout to denote integers divided by $n$, but in this case $n=1$ and so $p$ ranges over the integers. Substituting equation \eqref{finalTResult} into equation \eqref{eq:ooprimeprime}, we find
\begin{equation}\label{areaFinal}
	\replicaoperator \frac{-n}{2}\braket{{\mathcal{O}\mathcal{O}}}_{n}'' = \frac{1}{2\pi} \sum_{i,j,m} \int \frac{dr \, dr'}{(rr')^{2}} f^{(m-2)}_{ij}(r) f^{(-m-2)}_{ji}(r') e^{-\pi(K_{i}+K_{j})}\frac{\sinh{\pi\alpha_{ij}}}{im-\alpha_{ij}}.
\end{equation}

\subsection{Evaluation of Multi-Sheet Correlator}\label{sec-t2}
We now turn to the second term in \eqref{discreteO2ptFunction},
\begin{align}\label{secondTermAlone}
	\frac{1}{2} \replicaoperator \ev{{\left(\sum_{k=0}^{n-1}\mathcal{O}^{(k)}\right)^2}}''_n .
\end{align}
The analytic continuation of this term to real \(n\) will turn out to be much more challenging than that of the first term of \eqref{discreteO2ptFunction}, because \(n\) appears in the upper summation limit. 

Using \eqref{eq-Oexpr}, can write the sum over replicas in \eqref{secondTermAlone} as follows:
\be\label{entropyTwoPoint}
\ev{{\left(\sum_{k=0}^{n-1}\mathcal{O}^{(k)}\right)^2}}_n=\ev{{\left(\sum_{i,j} \int_0^{2\pi n} dr \, d\theta\, f_{ij}(r,\theta) \partial\Phi(r,\theta;\lambda) \otimes E_{ij}(\theta) \right)^2}}_n.
\ee
This equality comes from interpreting $\mathcal O^{(k)}$ as $\mathcal O$ inserted on the $(k+1)$th replica sheet (see \eqref{eq-Okdef}). Summing over sheets and integrating \(\theta \in [0,2\pi]\) on each one is equivalent to just integrating \(\theta \in [0,2\pi n]\), which covers the entire replicated manifold. The definition of $\partial \Phi$ for angles greater than $2\pi$ is given by the the Heisenberg evolution rule, the right hand side of \eqref{phases}. The field is still holomorphic, but it would be misleading to write it as a function of $re^{i\theta}$ since it is not periodic in $\theta$ with period $2\pi$.

Because the $f_{ij}(r,\theta)$ are not dynamical, they should be identical on each sheet. In the Fourier representation as in \eqref{fourierf'sn=1}, this means keeping the Fourier coefficients fixed and keeping the $m$ parameters integer. Thus we have
\begin{align}\label{entropy1}
	\frac{1}{2} \replicaoperator \ev{{\left(\sum_{k=0}^{n-1}\mathcal{O}^{(k)}\right)^2}}''_n 
	= 
	\replicaoperator \frac{1}{2(2\pi)^{2}} 
	\sum_{\substack{i,j,i',j'\\ m,m'}} 
	\int & dr \, dr' \, d\theta \, d\theta' \,
	f_{ij}^{(m)}(r) f_{i'j'}^{(m')}(r') e^{-im\theta} e^{-im'\theta'} \nonumber\\[-1em]
	& \times \braket{\partial\Phi(r,\theta) \partial\Phi(r',\theta')}_n'' \braket{E_{ij}(\theta) E_{i'j'}(\theta')}_n~.
\end{align}

The CFT two point function is calculated in Appendix \ref{app:fieldTheory}:
\begin{align}
	\braket{\partial \Phi(z) \partial \Phi(w)}''_{n} &= \frac{1}{n (zw)^{2}} \sum_{|q|<1} \sign(q) q(q^2 - 1) \left(\frac{w}{z}\right)^{q}\\
	&= \frac{1}{n(r r')^{2}} \sum_{|q|<1} \sign(q) P(q,r,r') e^{i\theta(-q-2)} e^{i\theta'(q-2)}\label{relevantCFTtwopointfunction}
\end{align}
where \(q\) takes values in the integers divided by \(n\), and
\be\label{P(q)def}
	P(q,r,r')  \equiv q(q^2 - 1) \left(\frac{r'}{r}\right)^{q}~.
\ee
When $n=1$ there are no nonzero terms in the sum, but when $n>1$ the answer is nonzero. For future convenience, we separated the parts which depend on \(\theta\) from those that do not.

The auxiliary system two point function is calculated in Appendix \ref{app:aux}:
\begin{align}
	\braket{{ E_{ij}(\theta)E_{i'j'}(\theta')}}_n = \delta_{ij'}\delta_{ji'} e^{-2\pi n K_{i}} \frac{1}{\pi n \tilde Z_n^{\rm aux}} \sum_{p} e^{-ip(\theta-\theta')} \frac{\sinh{n \pi \alpha_{ij}}}{ip+\alpha_{ij}}e^{n \pi\alpha_{ij}}~,
\end{align}
where \(p\) is also an integer divided by \(n\) and $\tilde Z_n^{\rm aux} \equiv \trace\left[e^{-2\pi n K^{(0)}_{\text{aux}}}\right]$ is a normalization factor. Substituting this equation as well as \eqref{relevantCFTtwopointfunction} into \eqref{entropy1} gives
\begin{align}
	\replicaoperator \frac{1}{n^{2}(2\pi)^{3}\tilde Z_n^{\rm aux}} \sum_{\substack{i,j,p\\ m,m'}} \int \frac{dr \, dr' \, d\theta \, d\theta' }{(rr')^{2}} f_{ij}^{(m)}(r) f_{ji}^{(m')}(r') e^{i\theta(-q-p-2-m)} e^{i\theta'(q+p-2-m')} \nonumber\\[-1em]
	\times \frac{\sinh{\pi n \alpha_{ij}}}{ip+\alpha_{ij}} e^{-\pi n(K_{i}+K_{j})} \sum_{|q|<1} \sign(q) P(q,r,r')~.
\end{align}

The angle integrations give Kronecker deltas multiplied by \(2\pi n\). The result is 
\begin{align}
	& \replicaoperator \frac{i}{2\pi \tilde Z_n^{\rm aux}} \sum_{i,j,m} \int \frac{dr \, dr'}{(rr')^{2}} f_{ij}^{(m-2)}(r) f_{ji}^{(-m-2)}(r') \sinh{\pi n \alpha_{ij}} e^{-\pi n(K_{i}+K_{j})} \left[\sum_{|q|<1} \frac{\sign(q) P(q,r,r')}{q+m+i\alpha_{ij}}\right] \nonumber\\
	&= \frac{i}{2\pi} \sum_{i,j,m} \int \frac{dr \, dr'}{(rr')^{2}} f_{ij}^{(m-2)}(r) f_{ji}^{(-m-2)}(r') \sinh{\pi \alpha_{ij}} e^{-\pi (K_{i}+K_{j})}\ \replicaoperator \left[\sum_{|q|<1} \frac{\sign(q) P(q,r,r')}{q+m+i\alpha_{ij}}\right] ~.\label{EntropyBeforeAnalyticCont}
\end{align}

In going to the last line, we used the fact that the sum in brackets vanishes when \(n=1\) and that, for any two functions \(f({n}), g({n})\) such that \(f({1})\) and \(\left[\frac{d}{dn}f(n)\right]_{n=1}\) are finite and \(g({1}) = 0\), the following relation holds:
\begin{align}\label{Rfg}
	\replicaoperator \left({f({n}) g({n})}\right) = f(1) \, \replicaoperator g(n)~.
\end{align}

We now turn to the analytic continuation and application of \(\replicaoperator\) on the term in brackets in \eqref{EntropyBeforeAnalyticCont}. We will take care of the awkward ${\rm sign}(q)$ by writing the $q$-dependent part of the sum as two sums with positive argument. We will suppress the \((r,r')\) dependence for the rest of the calculation:
\be
	\sum_{|q|<1} \frac{{\rm sign}(q)P(q)}{q+m+i\alpha_{ij}} = \sum_{0<q<1}  \frac{P(q)}{q+m+i\alpha_{ij}} + \frac{P(-q)}{q-m-i\alpha_{ij}}~.
\ee
Now we write $q=k/n$ to turn this into a sum over integers:
\be
 \sum_{0<q<1} \left( \frac{P(q)}{q+m+i\alpha_{ij}} + \frac{P(-q)}{q-m-i\alpha_{ij}}\right) = \sum_{k=1}^{n-1}\left( \frac{P(\frac{k}{n})}{\frac{k}{n}+m+i\alpha_{ij}} + \frac{P(-\frac{k}{n})}{\frac{k}{n}-m-i\alpha_{ij}}\right)~.
 \ee
In the next section we will see how to evaluate and analytically continue such sums quite generally.

\subsection{Analytic Continuation}\label{app:analyticCont}
We need to evaluate
\begin{align}\label{thingToAnalyticallyContinue}
	\replicaoperator \sum_{k=1}^{n-1} \left( \frac{P(\frac{k}{n})}{\frac{k}{n}-z} + \frac{P(-\frac{k}{n})}{\frac{k}{n}+z} \right)~,
\end{align}
where \(P(z)\) is given by \eqref{relevantCFTtwopointfunction}. However, for the remainder of this section, we will consider \(P(z)\) to be an arbitrary analytic function whose functional form is independent of \(n\). We will specialize to the form given by \eqref{relevantCFTtwopointfunction} in section \ref{putItTogether}.

We start by writing the sum in \eqref{thingToAnalyticallyContinue} as
\begin{align}\label{expandedAnalyticContSum}
	\replicaoperator \sum_{k=1}^{n-1} \left( \frac{P(\frac{k}{n})-P(z)}{\frac{k}{n}-z} + \frac{P(-\frac{k}{n})-P(z)}{\frac{k}{n}+z} \right) + \replicaoperator\sum_{k=1}^{n-1} \left( \frac{P(z)}{\frac{k}{n}-z} + \frac{P(z)}{\frac{k}{n}+z} \right)
\end{align}
and then we evaluate the terms separately. Consider the first term in the first set of parenthesis. Because \(P(z)\) is analytic, we can expand it in a power series with positive powers of \(z\): \(P(z) = \sum_{r=0}^{\infty} a_{r} z^{r}\). This gives
\begin{align}
	\replicaoperator \sum_{k=1}^{n-1} \sum_{r=1}^{\infty} a_{r} \frac{(\frac{k}{n})^{r}-z^{r}}{\frac{k}{n}-z}~.
\end{align}
We can simplify the fraction using polynomial division; for \(r \geq 1\),
\begin{align}
	\frac{(\frac{k}{n})^{r}-z^{r}}{\frac{k}{n}-z} = \sum_{s=0}^{r-1} z^{r-s-1}\left(\frac{k}{n}\right)^{s}~,
\end{align}
which means the first term in the first set of parenthesis in \eqref{expandedAnalyticContSum} is
\begin{align}
	\replicaoperator \sum_{k=1}^{n-1} \frac{P(\frac{k}{n})-P(z)}{\frac{k}{n} - z} = \sum_{r=1}^{\infty} \sum_{s=0}^{r-1}a_{r} z^{r-s-1} \replicaoperator\sum_{k=1}^{n-1} \left(\frac{k}{n}\right)^{s}~.
\end{align}
The advantage of writing it this way is that it isolates the \(n\) dependence into something which can be easily analytically continued. First, recall that overall factors of powers of \(n\) don't matter if the expression they multiply vanishes at \(n=1\), as in \eqref{Rfg}. Next, note that the resulting expression is actually a polynomial in \(n\). It can be expressed this way using \emph{Faulhaber's formula}:
\begin{align}
	\sum_{k=1}^{n-1}k^{s} = \frac{1}{s+1} \sum_{j=0}^{s} (-1)^{j} \binom{s+1}{j} B_{j} (n-1)^{s-j+1}~,
\end{align}
where \(B_{s}\) is the j-th Bernoulli number in the convention that \(B_{1} = -1/2\). This makes application of \(\replicaoperator\) straightforward:
\begin{align}
	\replicaoperator\sum_{k=1}^{n-1} \left(\frac{k}{n}\right)^{s} = \replicaoperator\sum_{k=1}^{n-1} k^{s} = -(-1)^{s} B_{s}~.
\end{align}
Thus for the first term in \eqref{expandedAnalyticContSum} we have
\begin{align}
	\replicaoperator \sum_{k=1}^{n-1} \frac{P(\frac{k}{n})-P(z)}{\frac{k}{n} - z} = -\sum_{r=1}^{\infty} \sum_{s=0}^{r-1}a_{r} z^{r-s-1} (-1)^{s} B_{s}~.
\end{align}
The second term follows completely analogously:
\begin{align}
	\replicaoperator \sum_{k=1}^{n-1}  \frac{P(-\frac{k}{n})-P(z)}{\frac{k}{n}+z} =  \sum_{r=1}^{\infty} \sum_{s=0}^{r-1} a_{r} z^{r-s-1} B_{s}~.
\end{align}
Combining these results, the first set of large parenthesis in \eqref{expandedAnalyticContSum} is
\begin{align}
	- \sum_{r=1}^{\infty} \sum_{s=0}^{r-1} a_{r} z^{r-s-1} B_{s} \left[ (-1)^{s} - 1\right]~.
\end{align}
For even \(s\) this is zero. For odd \(s>1\), \(B_{s} = 0\), and so only \(s=1\) can contribute. Substituting \(B_{1}=-1/2\) gives
\begin{align}\label{firstParenthesisResult}
	-\frac{P(z)}{z^{2}} + \frac{a_{1}}{z} + \frac{a_{0}}{z^{2}}~.
\end{align}

We now turn to the second set of parenthesis in \eqref{expandedAnalyticContSum}. These two terms can be evaluated simultaneously. First, we can multiply through by \(n/n\) to give an overall factor of \(n\) (which is irrelevant) and convert the denominators to \(k-zn\) and \(k+zn\). We also pull \(P(z)\) through \(\replicaoperator\) because it is independent of \(n\):
\begin{align}
	P(z) \replicaoperator\sum_{k=1}^{n-1} \left( \frac{1}{k-zn} + \frac{1}{k+zn} \right)~.
\end{align}
This sum can be evaluated in terms of the digamma function \(\psi^{(0)}(w)\), which is defined in terms of the Gamma function \(\Gamma(w)\):
\begin{align}
	\psi^{(0)}(w) \equiv \frac{\Gamma'(w)}{\Gamma(w)} = -\gamma + \sum_{k=0}^{\infty} \left( \frac{1}{k+1} - \frac{1}{k+w} \right)~.
\end{align}
By manipulating the sum, one can show
\begin{align}
	\sum_{k=1}^{n-1} \frac{1}{k-w} = \psi^{(0)}(n-w) - \psi^{(0)}(1-w)~.
\end{align}¥
Thus the second set of parenthesis in \eqref{expandedAnalyticContSum} is equal to
\begin{align}\label{secondParenthesis}
	 P(z) \replicaoperator \left[ \psi^{(0)}(n-zn) - \psi^{(0)}(1-zn) + \psi^{(0)}(n+zn) - \psi^{(0)}(1+zn)\right]~.
\end{align}

We cannot naively apply \(\replicaoperator\) yet. We first have to select the correct analytic continuation to real positive \(n\) from the many possible analytic continuations of integer \(n\) data. This is known to be a challenging problem in general.\footnote{See Ref.~\cite{Fau14} for a recent discussion of the difficulties of the analytic continuation. Ref.~\cite{Fau14} also contains another method for computing the entropy perturbatively that does not rely on the replica trick. Such a method avoids the need to analytically continue, and applying it to the present calculation would serve as a check of our analytic continuation prescription. We leave that check to future work.} Nevertheless, in our context the correct analytic continuation prescription is clear. 

The digamma function has poles in the complex plane at zero and all negative real integers. Recall that we are ultimately interested in plugging in \(z_{m} \equiv -m-i\alpha_{ij}\). Thus if we are not careful, for certain values of \(m\), the digamma functions in \eqref{secondParenthesis} will blow up when \(\alpha_{ij} \to 0\) near \(n=1\). On the other hand, on physical grounds we expect our result to be perfectly well-behaved when \(\alpha_{ij} \to 0\), which simply corresponds to a degeneracy in the auxiliary system. The way we avoid the poles of the digamma function near \(n=1\) when \(\alpha_{ij} \to 0\) is by using the reflection formula
\begin{align}\label{psiRefl}
	\psi^{(0)}(1-w) = \psi^{(0)}(w) + \pi \cot{\pi w}~,
\end{align}
which produces different analytic continuations given the same integer data. These observations lead to the following prescription: \emph{for each value of \(m\), use the reflection formula \eqref{psiRefl} to avoid the poles of the digamma function near \(n=1\) as \(\alpha_{ij} \to 0\).}

As an example, consider the term \(\psi^{(0)}(1-z_{m} n) = \psi^{(0)}(1+mn + \alpha_{ij} n)\) in \eqref{secondParenthesis}. When \(\alpha_{ij} = 0\), this has a pole when \(nm \leq 1\). Thus for a given \(m \leq 1\), we cannot expect to have a smooth \(n\)-derivative at \(n=1\). The resolution is to use \eqref{psiRefl} to get
\begin{align}
	\psi^{(0)}(1+mn + \alpha_{ij} n) &= \psi^{(0)}(-mn-\alpha_{ij}) - \pi \cot{\pi(mn+\alpha_{ij})} \\
	&= \psi^{(0)}(-mn-\alpha_{ij}) - \pi \cot{\pi\alpha_{ij}}~,
\end{align}
where the last equality is only true for integer \(n\). The remaining digamma term is now free of poles for \(mn \leq 1\), which is precisely when there was a problem before the application of the reflection formula, and \(\replicaoperator\) can now be easily applied. This example illustrates how the correct analytic continuation depends on the value of \(m\). We must apply this reasoning separately to each term in \eqref{secondParenthesis}. After applying this procedure to each digamma function as needed to avoid the poles, it will turn out that all of the extra cotangent terms cancel against each other.

\begin{figure}\label{fig:analyticCont}
\centering
	\includegraphics[width=.9\textwidth]{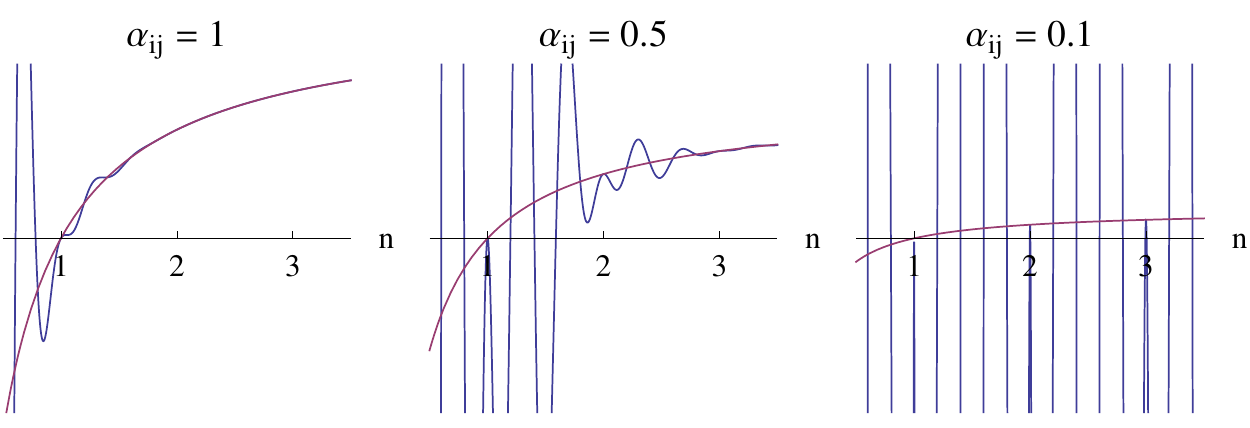}
	\caption{Sample plots of the imaginary part (the real part is qualitatively identical) of the na\"ive bracketed digamma expression in \eqref{secondParenthesis} and the one in \eqref{eq-continuation} obtained from analytic continuation with \(z=-m-i\alpha_{ij}\) for \(m=3\) and various values of \(\alpha_{ij}\). The oscillating curves are \eqref{secondParenthesis}, while the smooth curves are the result of applying the specified analytic continuation prescription to that expression, resulting in \eqref{eq-continuation}.}
\end{figure}

There is another way to motivate this prescription. Even for small but finite \(\alpha_{ij}\), the analytic continuations picked out by our prescription can be seen to be qualitatively better than the one obtained by using \eqref{secondParenthesis} directly, as illustrated in Figure \ref{fig:analyticCont}. Notice that while both curves match for integer \(n\), the curve obtained by applying the prescription outlined above is the only one which smoothly interpolates between the integers. The oscillations of the ``wrong'' curves get larger and larger as \(\alpha_{ij}\) is reduced or \(m\) is increased.

Applying our prescription to \eqref{secondParenthesis}, there are three expressions depending on the value of \(m\). We are focussing on the quantity in brackets in \eqref{secondParenthesis}: 
\begin{align}\label{eq-continuation}
	\begin{cases}
		\psi^{(0)}(1-n-nz_{m}) - \psi^{(0)}(-nz_{m}) + \psi^{(0)}(n-nz_{m}) - \psi^{(0)}(1-nz_{m}) & m>0 \\
		\psi^{(0)}(n+nz_{m}) - \psi^{(0)}(1+nz_{m}) + \psi^{(0)}(n-nz_{m}) - \psi^{(0)}(1-nz_{m}) & m=0 \\
		\psi^{(0)}(n+nz_{m}) - \psi^{(0)}(1+nz_{m}) + \psi^{(0)}(1-n+nz_{m}) - \psi^{(0)}(nz_{m}) & m<0
	\end{cases}
\end{align}
Now we are ready to apply \(\replicaoperator\). The digammas \(\psi^{(0)}(w)\) will turn into polygammas \(\psi^{(1)}(w) \equiv \frac{d}{dw} \psi^{(0)}(w)\), which obey the recurrence relation 
\begin{align}
	\psi^{(1)}(w+1) = \psi^{(1)}(w) - \frac{1}{w^{2}}~.
\end{align}
This recurrence relation simplifies the result for \(m>0\) and \(m<0\) while the recurrence relation along with the reflection formula simplifies the result for \(m=0\). The result for the second set of parenthesis in \eqref{expandedAnalyticContSum} with \(z=z_{m}\) is
\begin{align}\label{secondParenthesisResult}
	\frac{P(z_{m})}{z_{m}^{2}} + \delta(m) P(z_{m}) \frac{\pi^{2}}{\sinh^{2}{\pi\alpha_{ij}}}~.
\end{align}

We are now ready to give the final expression for \eqref{thingToAnalyticallyContinue}. Adding \eqref{firstParenthesisResult} with \(z=z_{m}\) and \eqref{secondParenthesisResult} we find
\begin{align}\label{generalAnalyticContResult}
	\replicaoperator \sum_{k=1}^{n-1} \left( \frac{P(\frac{k}{n})}{\frac{k}{n}-z_{m}} + \frac{P(-\frac{k}{n})}{\frac{k}{n}+z_{m}} \right) = \frac{a_{1}}{z_{m}} + \frac{a_{0}}{z_{m}^{2}} + \delta(m) P(-i\alpha_{ij}) \frac{\pi^{2}}{\sinh^{2}{\pi\alpha_{ij}}}
\end{align}
for arbitrary analytic \(P(z_{m})\).

\subsection{Completing the Proof}\label{putItTogether}
Now we specialize to the form of \(P(z)\) needed for our calculation which came from the particular \(\braket{\partial\Phi \partial\Phi}''\) two-point function we were computing (\eqref{relevantCFTtwopointfunction} and \eqref{P(q)def}):
\begin{align}
	P(z) = z(z^{2}-1) e^{z\log{(r'/r)}}~.
\end{align}
Thus \(a_{0}=0\), and \(a_{1}=-1\). Using \eqref{generalAnalyticContResult} gives 
\begin{align}
	&\replicaoperator \sum_{k=1}^{n-1} \left( \frac{P(\frac{k}{n})}{\frac{k}{n}-z_{m}} + \frac{P(-\frac{k}{n})}{\frac{k}{n}+z_{m}} \right) = 
	&\frac{i}{im-\alpha_{ij}} + \delta(m)\frac{i\pi^{2}}{\sinh^{2}{\pi\alpha_{ij}}}\alpha_{ij}(\alpha_{ij}^{2}+1)\left(\frac{r'}{r}\right)^{-i\alpha_{ij}}~.
\end{align}
Plugging this into \eqref{EntropyBeforeAnalyticCont} and plugging that into \eqref{entropyTwoPoint} gives the term from \eqref{discreteO2ptFunction} that we have been focussing on in this section:
\begin{align}\label{finalSResult}
	\replicaoperator \frac{1}{2}\ev{{\left(\sum_{k=0}^{n-1}\mathcal{O}^{(k)}\right)^2}}''_n = &\frac{-1}{2\pi} \sum_{i,j,m}\int \frac{dr dr'}{(r r')^{2}} f^{(m-2)}_{ij}(r) f^{(-m-2)}_{ji}(r') \sinh{\pi \alpha_{ij}} e^{-\pi(K_{i}+K_{j})}\nonumber\\
	& \times \left[\frac{1}{im-\alpha_{ij}} +\delta(m)\frac{\alpha_{ij}}{\sinh^{2}{\pi\alpha_{ij}}} \pi^{2}(\alpha_{ij}^{2}+1) \left(\frac{r'}{r}\right)^{-i\alpha_{ij}} \right] ~.
\end{align}

Notice that the first term in this expression exactly cancels the contribution to ${S^{(2)\dprime}}$ coming from the first term in \eqref{discreteO2ptFunction}, presented in \eqref{areaFinal}. We now consider the second term, and define the manifestly positive quantity \(M_{ij} \equiv e^{-\pi(K_{i}+K_{j})}\pi^{2}(\alpha_{ij}^{2}+1)\) to clean up the notation. Then we have
\begin{align}\label{SgenRaw}
	{S^{(2)\dprime}}= \frac{-1}{2\pi} \sum_{i,j} \int \frac{dr dr'}{(r r')^{2}} & f^{(-2)}_{ij}(r) f^{(-2)}_{ji}(r') \left(\frac{r'}{r}\right)^{-i\alpha_{ij}} \frac{\alpha_{ij}}{\sinh{\pi\alpha_{ij}}} M_{ij}~.
\end{align}
The integrals over \(r,r'\) factorize, giving
\begin{align}\label{factorized}
	{S^{(2)\dprime}}= \frac{-1}{2\pi} \sum_{i,j} \left[ \int_{0}^{\infty} dr\ r^{i\alpha_{ij}-1} f^{(-2)}_{ij}(r) \right] \left[ \int_{0}^{\infty} dr\ r^{-i\alpha_{ij}-1} f^{(-2)}_{ji}(r) \right] \frac{\alpha_{ij}}{\sinh{\pi\alpha_{ij}}} M_{ij}~.
\end{align}
Recall the constraint on the test functions derived previously by requiring the density matrix be Hermitian (equation \eqref{hermThetaConstraint}): \(f_{ij}(r,\theta) = f_{ji}(r,2\pi -\theta)^*\). In Fourier space, this implies \(f^{(m)}_{ji}(r) = f^{(m)}_{ij}(r)^{*}\). Inserting this into \eqref{factorized} we see that the factors in brackets are complex-conjugates of each other. Furthermore, because \(\sinh{\pi\alpha_{ij}}\) always has the same sign as \(\alpha_{ij}\), the overall sign of the entire term is negative and so we find
\begin{align}
	{S^{(2)\dprime}}\leq 0~.
\end{align}
As discussed after \eqref{QNECandS2}, this proves the QNEC.

\section{Extension to $D=2$, Higher Spin, and Interactions}\label{sec-extensions}

In $D = 2$, there are no transverse directions, and so it is not possible to use the fact that the state is very close to the vacuum.  Nevertheless, once one has proven the QNEC for a free scalar field in $D > 2$, one can use dimensional reduction to prove it for free scalar fields in $D=2$. Let $\Phi(z,y)$ be the chiral scalar on $N$ in $D>2$, where $y$ labels the $D-2$ transverse coordinates. One can isolate a single transverse mode by integrating $\Phi(z,y)$ against a real transverse wavefunction, and this defines an effective two-dimensional field:
\be
\Phi_{2D}(z) \equiv \int dy\, \psi(y) \Phi(z,y)~,
\ee
where $\psi$ is normalized such that $\int \psi^2 = 1$. Correlation functions of $\Phi_{2D}$ and its derivatives exactly match those of a two-dimensional chiral scalar, and so our dimensional reduction is defined by the subspace of the $D$-dimensional theory obtained by acting on the vacuum with $\Phi_{2D}$. In any such state, one can integrate the $D$-dimensional QNEC along the transverse direction to find
\be
\int dy\, \braket{T_{kk}(y)} \geq \frac{1}{2\pi}\int dy\, \frac{\delta^2 S_{\rm out}}{\delta \lambda(y)^2}~.
\ee
Here we have suppressed the value of the affine parameter as a function of the transverse direction. The effective two-dimensional change in the entropy is defined by considering a total variation in all of the generators which is uniform in the transverse direction. For such a variation we have
\be
S_{2D}'' = \int dy\,dy'\,\frac{\delta^2 S_{\rm out}}{\delta \lambda(y)\delta \lambda(y')} \leq  \int dy\, \frac{\delta^2 S_{\rm out}}{\delta \lambda(y)^2}~,
\ee
where the the inequality comes from applying strong subadditivity to the off-diagonal second derivatives~\cite{BouFis15a}. The two-dimensional energy momentum tensor is defined in terms of the normal ordered product of the two-dimensional fields, $T_{2D} = :\partial \Phi_{2D}\partial \Phi_{2D}:$. However, using Wick's theorem one can easily check that $T_{2D}$ acts on the dimensionally reduced theory in the same way as the integrated $D$-dimensional $T_{kk}$:
\be
\braket{T_{2D}(w) \Phi_{2D}(z_1)\cdots \Phi_{2D}(z_n)} = \int dy\,\braket{T_{kk}(w,y) \Phi_{2D}(z_1)\cdots \Phi_{2D}(z_n)}~.
\ee 
Therefore the QNEC holds for a free scalar field in two dimensions:
\be
\braket{T_{2D}} = \int dy\, \braket{T_{kk}(y)} \geq \frac{1}{2\pi}\int dy\, \frac{\delta^2 S_{\rm out}}{\delta \lambda(y)^2} \geq \frac{1}{2\pi}S_{2D}''.
\ee

The extension to bosonic fields with spin is trivial, as these simply reduce on $N$ to multiple copies of the 1+1 chiral scalar CFT, one for each polarization.  These facts are reviewed in \cite{Wall11}.  Similarly, fermionic fields reduce to the chiral 1+1 fermion CFT; we expect that there is a similar proof in this case.

Astute readers may have noticed that the mass term of the higher dimensional field theory plays no role in our analysis.  Since it does not contribute to the commutation relations on $N$ or to $T_{kk}$, it plays no role in our analysis.  Regardless of whether the $D$ dimensional theory has a mass, the $1+1$ chiral theory is massless.  In a sense, null surface quantization is a UV limit of the field theory.  One might therefore expect that the addition of interactions with positive mass dimension (superrenormalizable couplings) will also not change the algebra of observables on $N$.  So long as this is the case, the extension to theories with superrenormalizable interactions is trivial.

One argument that superrenormalizable interactions are innocuous proceeds in two stages \cite{Wall11}.  First, one considers the direct effects of adding interaction terms to the Lagrangian; for example a scalar field potential $V(\phi)$.  So long as these interaction terms contain no derivatives (or are Yang-Mills couplings), they do not contribute to the commutation relations of fields restricted to the null surface, or to $T_{kk}$.  (So far, the interaction could be of any scaling dimension, so long as one avoids derivative couplings.)  

Next, one considers loop corrections due to renormalization.  In the case of a marginally renormalizable, or nonrenormalizable theory, these loop corrections normally require the addition of counterterms containing derivatives (for example, field strength renormalization), spoiling the null surface formulation.  On the other hand, in a superrenormalizable theory, only couplings with positive mass dimension require counterterms.  For a standard QFT consisting of scalars, spinors, and/or gauge fields, none of these superrenormalizable interactions include the possibility derivative couplings.  Thus one expects that loop corrections do not spoil the algebra of observables on the null surface.  However, superrenormalizable theories are difficult to construct except when $D < 4$.  (For example, the $\phi^3$ theory is superenormalizable in $D < 6$, but is unstable.)

It is an open question whether the QNEC is valid for non-Gaussian $D = 2$ CFT's in states besides conformal vacua, or more generally for QFT's in any dimension which flow to a nontrivial UV fixed point.\footnote{In more than 2 dimensions, interacting CFTs appear to have no nontrivial observables on the horizon\cite{Wall11,BCFM2}, so the current proof cannot be extended to this situation.}  Nor have we carefully considered the effects of making the scalar field noncompact.  QCD in $D = 4$ is a borderline case; the coupling flows to zero, but slowly enough that there is an infinite field strength renormalization.  Strictly speaking this makes null surface quantization invalid, yet it is still a useful numerical technique for studying hadron physics \cite{Burkardt:1995ct}.  However, we conjecture that the QNEC will be true in every QFT satisfying reasonable axioms.

\section*{Acknowledgements}
It is a pleasure to thank C.~Akers, E.~Bianchi, N.~Engelhardt, T.~Jacobson, J.~Maldacena, D.~Marolf, and D.~Simmons-Duffin for discussions. 
The work of RB, ZF, JK, and SL is supported in part by the Berkeley Center for Theoretical Physics, by the National Science Foundation (award numbers 1214644, 1316783, and 1521446), by fqxi grant RFP3-1323, and by the US Department of Energy under Contract DE-AC02-05CH11231. The work of AW is supported in part by NSF grant PHY-1314311 and the Institute for Advanced Study.

\appendix
\section{Correlation Functions}

\subsection{Scalar Field}\label{app:fieldTheory}
The chiral scalar operator \(\partial\Phi(z)\) is a conformal primary of dimension \((h,\bar{h}) = (1,0)\). Its two point function on the Euclidean plane is fixed by conformal symmetry up to an overall constant. We will take the following normalization:
\begin{align}
	\braket{\partial \Phi(z) \partial \Phi(w)} = \frac{-1}{(z-w)^{2}}~.
\end{align}
The two point function on the \(n\)-sheeted replicated manifold is obtained by application of the conformal transformation \(z \to z^{n}\):
\begin{align}
	\braket{\partial \Phi(z) \partial \Phi(w)}_{n} = \frac{-1}{n^{2} z w} \frac{(z w)^{1/n}}{(z^{1/n}-w^{1/n})^{2}}~.
\end{align}

The second-derivative of this two point function under translations of the holomorphic coordinate, evaluated at $\lambda=0$, is defined by
\begin{align}
	 \braket{\partial \Phi(z-\lambda) \partial \Phi(w-\lambda)}_{n}'' = \braket{\partial^{3} \Phi(z) \partial \Phi(w)}_{n} + \braket{\partial \Phi(z) \partial^{3} \Phi(w)}_{n} + 2\braket{\partial^{2} \Phi(z) \partial^{2} \Phi(w)}_{n}~.
\end{align}
One can show that this combination of correlation functions can be written as 
\begin{align}
	\frac{1}{n (zw)^{2}} \sum_{|q|<1} \sign(q) q(q^2 - 1) \left(\frac{w}{z}\right)^{q}~,
\end{align}
where \(q\) is an integer divided by \(n\). Notice that this implies that the sum vanishes for $n=1$, as required by translation invariance. 

Our convention for the only nonzero component of the stress tensor for the holomorphic sector of the theory is
\begin{align}
	T(z) = -2\pi T_{zz}(z) = -\frac{1}{2} :\partial\Phi(z)\partial\Phi(z):~,
\end{align}
where \(:AB:\) denotes the normal-ordered product. Thus using Wick's theorem we have
\begin{align}\label{Tphiphi}
	\braket{\partial\Phi(z) \partial\Phi(w)T(0)} = \frac{-1}{(zw)^{2}}~.
\end{align}

\subsection{Auxiliary System}\label{app:aux}

In this appendix we will evaluate the \(\theta\)-ordered correlation functions of the auxiliary system,
\be
\braket{{E_{ij}(\theta)E_{i'j'}(\theta')}}_n = \frac{\trace\left[e^{-2\pi n K_{\text{aux}}}\thetaorder{E_{ij}(\theta)E_{i'j'}(\theta')}\right]}{\trace\left[e^{-2\pi n K_{\text{aux}}}\right]}~.
\ee
First, consider the case \(\theta>\theta'\):
\begin{align}\label{thermalExpect1}
	\trace\left[e^{-2\pi n K_{\text{aux}}}E_{ij}(\theta)E_{i'j'}(\theta')\right] = e^{-2\pi n K_{i}}e^{(\theta-\theta')\alpha_{ij}}\delta_{ij'}\delta_{ji'}~,
\end{align}
where $\alpha_{ij} \equiv K_i - K_j$ is the difference in two of the eigenvalues of $K_{\rm aux}$. For \(\theta<\theta'\), we have the opposite ordering inside the expectation value, which gives
\begin{align}\label{thermalExpect2}
	\trace\left[e^{-2\pi n K_{\text{aux}}}E_{i'j'}(\theta')E_{ij}(\theta)\right] = e^{-2\pi n K_{i}}e^{(\theta-\theta'+2\pi n)\alpha_{ij}}\delta_{ij'}\delta_{ji'}
\end{align}
We will find it convenient to use the following complex exponential representation of \(e^{(\theta-\theta')\alpha_{ij}}\), valid for \(\theta-\theta' \in (0,2\pi n)\):
\begin{align}\label{complexExp}
	e^{(\theta-\theta')\alpha_{ij}} = \frac{1}{\pi n} \sum_{p} e^{-ip(\theta-\theta')} \frac{\sinh{n \pi \alpha_{ij}}}{ip+\alpha_{ij}}e^{n \pi\alpha_{ij}}.
\end{align}
Here $p$ is being summed over all rational numbers which are integers divided by $n$. This can be substituted directly into \eqref{thermalExpect1}. For the expectation value when \(\theta<\theta'\) given by \eqref{thermalExpect2}, we can take \(\theta-\theta'+2\pi n\) as our Fourier series variable instead of \(\theta-\theta'\), which also lies in \((0,2\pi n)\) in this case. This means we can substitute this into \eqref{complexExp}, giving the same complex exponential representation:
\begin{align}
	e^{(\theta-\theta'+2\pi n)\alpha_{ij}} = \frac{1}{\pi n} \sum_{p} e^{-ip(\theta-\theta')} \frac{\sinh{n \pi \alpha_{ij}}}{ip+\alpha_{ij}}e^{n \pi\alpha_{ij}}.
\end{align}
Collecting these results, the $\theta$-ordered correlation function in the auxiliary system is simply
\begin{align}\label{thermalExpect}
	\braket{{ E_{ij}(\theta)E_{i'j'}(\theta')}}_n = \delta_{ij'}\delta_{ji'} e^{-2\pi n K_{i}} \frac{1}{\pi n \tilde Z_n^{\rm aux}} \sum_{p} e^{-ip(\theta-\theta')} \frac{\sinh{n \pi \alpha_{ij}}}{ip+\alpha_{ij}}e^{n \pi\alpha_{ij}},
\end{align}
where
\be
\tilde Z_n^{\rm aux} \equiv \trace\left[e^{-2\pi n K_{\text{aux}}}\right].
\ee
Note that $\tilde Z_1^{\rm aux} =1$.

\bibliographystyle{utcaps}
\bibliography{all}

\end{document}